\documentclass[
]{ceurart}

\usepackage{listings}
\begin{document}

\copyrightyear{2026}
\copyrightclause{Copyright for this paper by its authors.
  Use permitted under Creative Commons License Attribution 4.0
  International (CC BY 4.0).}

\conference{CLEF 2026 Working Notes, 21 -- 24 September 2026, Jena, Germany}

\title{SciClaimSeekers at CheckThat! 2026: Retrieving Scientific Sources for Social Media Claims with LLM Reranking}

\title[mode=sub]{CheckThat! Lab at CLEF 2026}

\author[1]{Mohotarema Rashid}[%
orcid=0009-0003-2683-7191,
email=MohotaremaRashid@my.unt.edu,
]
\cormark[1]

\address[1]{Department of Information Science, University of North Texas, Denton, TX, United States}
\address[2]{Department of Computer Science and Engineering, University of North Texas, Denton, TX, United States}
\address[3]{Department of Data Science, University of North Texas, Denton, TX, United States}

\author[2]{Nansu Baniya}[%
orcid=0000-0001-7116-9338,
email=NansuBaniya@my.unt.edu,
]
\author[2]{Anirban Saha Anik}[%
orcid=0000-0002-7824-3702,
email= AnirbanSahaAnik@my.unt.edu,
]
\author[1]{Xiaoying Song}[%
orcid=0000-0001-9390-1155,
email=XiaoyingSong@my.unt.edu,
]

\author[3]{Lingzi Hong}[%
orcid=0000-0001-8412-8180,
email=Lingzi.Hong@unt.edu,
]

\cortext[1]{Corresponding author.}

\begin{abstract}
  Scientific claims often spread on social media faster than they can be verified, while posts rarely link to the original scholarly sources. To tackle this problem this paper presents system called SciClaimSeekers, a retrieval and reranking framework by combining BM25 and zero-shot multilingual E5 retrieval with Reciprocal Rank Fusion (k=60), followed by Qwen2.5-14B-Instruct pointwise reranking. The pipeline reaches 64.36\% MRR@5 on the English development set a 13.67-point jump over BM25 and 10.17 points over the unranked hybrid and 64.39\% on the official test set, in the CLEF-2026 CheckThat! Task 1 evaluation. Our experiment suggests that large pre-trained models, when combined into a careful pipeline, can be competitive with fine-tuned approaches on this task.
\end{abstract}

\begin{keywords}
  Large Language Model \sep
  Retrieval \sep
  Social Media \sep
  CLEF
\end{keywords}

\maketitle

\section{Introduction}

Since online misinformation spreads rapidly, tracing social media claims back to their original scientific sources is crucial for automated fact checking and verification based on evidence~\cite{muhammed2022,sager2025}. Furthermore, users in social media tend to favor information that aligns with their preexisting beliefs and this confirmation bias, combined with the rapid dissemination of scientific information, facilitates the widespread acceptance and propagation of unverified claims as factual information~\cite{altay2024,moravec2018}. Because the content of social media is solely user-generated and user-generated posts often paraphrase or loosely reference findings without standardized citations, making reliable identification of the original scientific sources difficult~\cite{zhang2021}. Furthermore, this task poses challenges due to the linguistic and structural mismatch between informal social media jargon and formal scientific literature~\cite{booth2025}.

\begin{figure}[ht]
    \centering
    \includegraphics[width=\linewidth]{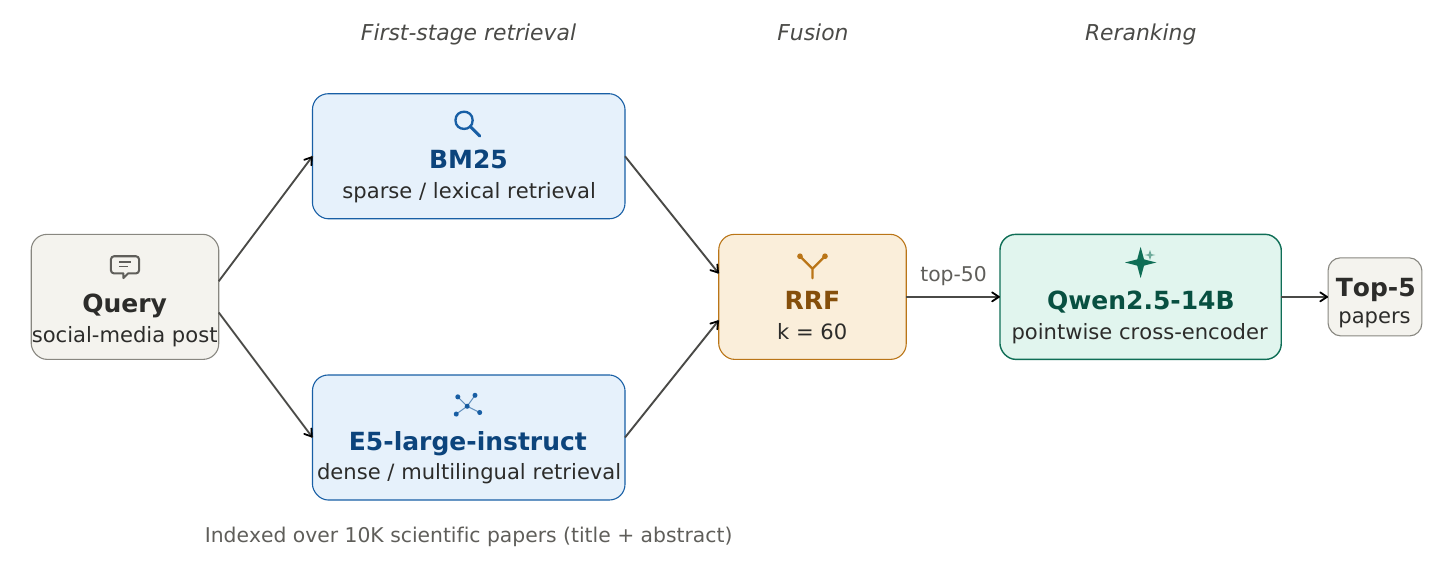}
    \caption{SCiClaimSeekers System Architecture}
    \label{fig:pipeline}
\end{figure}

To address this challenge our work proposed a combined reranking foundations integrating hybrid sparse dense retrieval with Large Language Model (LLM); Fig.~\ref{fig:pipeline} illustrates our three-stage retrieve-then-rerank architecture: two complementary first-stage retrievers operate in parallel on the indexed corpus, their ranked candidate lists are merged with Reciprocal Rank Fusion, and the fused candidate set is reranked by a LLM used as a pointwise cross-encoder.

In the above context, our contributions are as follows:

\begin{enumerate}
    \item We present a hybrid retrieve-then-rerank pipeline for the multilingual CheckThat! 2026 setting~\cite{10.1007/978-3-032-21321-1_43, clef-checkthat:2026-lncs}. Our submitted system ranks 11th out of 37 participating teams on the English test set for Task 1, Source Retrieval for Scientific Web Claims~\cite{clef-checkthat:2026:task1}, achieving 64.39\% MRR@5. The negligible 0.03 percentage-point gap from the development score of 64.36\% suggests that our tuning decisions transfer cleanly to held-out test data.
    \item We isolate the contribution of each pipeline stage and identify the LLM cross-encoder as the dominant source of gain approximately +10 MRR@5 points whether applied on top of BM25 or the hybrid first stage accounting for roughly three-quarters of the +13.67-point overall improvement over the BM25 baseline.
\end{enumerate}

\section{Related Work}

Our study is situated in intersection four major research streams I) automated fact checking II) scientific claim verification III) social media based scientific discourse retrieval and IV) LLM based reranking systems.

Early factchecking pipeline was shaped by FEVER benchmark~\cite{thorne2018} while SciFact extended the framework dedicated to scientific claims~\cite{wadden2020}. Later in this research stream ARSJoint, and RerrFact improved retrieval and verdict using transformer-based architectures~\cite{pradeep2023,rana2022}.

Though the rise of social media as a channel for scientific communication has created challenges for retrieving scholarly sources from implicit and noisy posts, early work in scientific claim verification rarely focused on noisy and implicit social media discourse. To shed light on this, SciTweets was developed for detecting scientific discourse from social media posts~\cite{hafid2022}. Building on this line of work, Hafid et al.~\cite{hafid2025disambiguation} formalized the task of disambiguating implicit scientific references on X, retrieving the original publications that social media posts implicitly refer to, and contributed the first ground-truth corpus and baselines for the task. Most prior systems adopted a retrieve-then-rerank architecture by combining sparse and dense retrieval~\cite{ashbaugh2025,lee2023,ren2023,yang2019}.

Traditional retrieval system like BM25 which based on lexical matching remains a strong classical sparse retrieval baseline due to its probabilistic ranking formulation~\cite{robertson2009}. In addition transformer-based dense retrievers such as DPR improved semantic matching for retrieval tasks~\cite{karpukhin2020}. Though both of the methods can combined into hybrid retrieval has become dominant as sparse retrievers effectively capture lexical signals such as named entities and numerical values, whereas dense retrievers better model semantic similarity and paraphrastic variation~\cite{khattab2020}.

In recent, reranking approaches increasingly use LLM model as cross encoders. Though early transformer based rankers such as monoBERT utilized reranking as relevance classification task~\cite{nogueira2020}, later studies use list wise strategy to improve ranking performance~\cite{pradeep2023}.

Recently LLM based closed rerankers such as RankGpt showed strong performance through instruction tuned model~\cite{sun2024}. However, open source alternatives also improved the retrieval task while maintaining competitive effectiveness~\cite{ma2024}.

In sum, prior studies showed substantial progress in scientific claim verification and retrieval, however challenges remain in linking noisy social media discourse with scholarly evidence. Our work addresses this gap by combining reranking foundations integrating hybrid sparse dense retrieval with LLM-based pointwise reranking for scientific source attribution in social-media environments. Because Pointwise remains promising due to its efficiency, scalability and lower sensitivity to input ordering bias~\cite{zhuang2024}.

\section{Methods}

\subsection{Pipeline overview}

We formulate the retrieval problem as follows; Given a social media post $q$ written in English the system must return a ranked list of five publications extracted from a fixed candidate pool $D$ of $|D| = 10{,}000$ scientific papers. Each candidate paper is represented by key, a title, an abstract and metadata fields like authors, venue, publication year.

Our pipeline decomposes the task into three computational sequential Phases.

In Phase 1, a sparse retriever based on BM25~\cite{robertson2009} and a dense retriever based on multilingual-e5-large-instruct~\cite{wang2024} operate in parallel and each independently returns its top-100 candidates per query.

In Phase 2, the two returned lists were merged using Reciprocal Rank Fusion~\cite{cormack2009} that returned an unified top-50 candidates set against each query.

In phase 3, each of the 50 candidates were scored by Qwen2.5-14B-Instruct~\cite{yang2024}. The scoring set up was conducted in a zero shot pointwise cross-encoder and the reranked settings produced top 5 candidates as final predictions.

\subsection{Data representation and indexing}

We represent each candidate paper concatenated by title and abstract, we ignored meta data field as it could introduce noise to the pipeline followed by prior literature~\cite{sager2025}. We indexed full corpus once at preprocessing time for both set of retrievals discussed later in this section.

\subsection{Sparse Retrieval}

We use Okapi BM25~\cite{robertson2009} with its standard parameter to work on retrieval that is based on lexical overlap between social media claim and title and abstract of the candidates. Though we note that lexical matching is insufficient this type of task since social media post contains online jargon and most of the cases the post paraphrase the scientific languages, yet BM25 remains a robust first-stage retriever for queries that include named entities, numerical values, and domain-specific terminology, which keeps the pipeline easy to reproduce~\cite{thakur2021}. To make this phase effective we preprocessed data by lowering the alphabetic characters, removing mentions and URL, normalizing hashtag, removing of punctuation, tokenizing of subwords, removing stopwords~\cite{sager2025,sennrich2016}.

\subsection{Dense Retrieval}

For dense retrieval we use \texttt{intfloat/multilingual-e5-large-instruct} retriever~\cite{wang2024} encoder which is based on built on XLM-RoBERTa. By following this model's convention, we differentiated queries from documents via instruction prefix, each post was tagged \texttt{"Instruct: Given a social media post that paraphrases a scientific finding, retrieve the scientific paper it refers to. \textbackslash nQuery: "}. In this setting we tokenized the input into 512 tokens and mean pooled over the final hidden states and we normalize L2 so that cosine similarity could be reduced to inner product~\cite{reimers2019}. We match query embeddings against FAISS index that returns top-100 documents per query.

\subsection{Reciprocal Rank Fusion}

In this stage we merge the two top-100 candidate lists using Reciprocal Rank Fusion~\cite{cormack2009}, which assigns scores to each document solely by its rank position in each input ranking.

\begin{equation}
\mathrm{RRF}(d) = \sum_{r \in R} \frac{1}{k + \mathrm{rank}_r(d)}
\end{equation}

where $R$ is the set of retrievers (BM25 and dense), $\mathrm{rank}_r(d)$ is the rank of $d$ in retriever $r$'s list ($\infty$ if absent), and $k$ is a smoothing constant. We use $k=60$ recommended by prior research~\cite{cormack2009}. We chose this depth based on the findings from the development set (discussed later in result section).

\subsection{LLM based point-wise reranking}

In this stage we use Qwen2.5-14B-Instruct~\cite{yang2024} as a setting of zero-shot pointwise cross-encode. This LLM is open-source serving, and 32K-token context window. We score each candidate independently rather than listwise~\cite{sun2024}.

For each (post, candidate) pair, we build a prompt instruct. The system message tells the model to act as a scientific source identifier and return an integer between 0 and 10, and the user message provides the post along with the candidate's title and abstract (Table~\ref{tab:prompt}). The relevance score were set to 0 to 10, then the Candidates are sorted by score, with ties broken using the Stage 2 RRF score, and the top 5 are returned.

\begin{table}[h]
    \centering
    \caption{Prompt for LLM to rerank pipeline.}
    \label{tab:prompt}
    \small
    \begin{tabular}{p{0.92\linewidth}}
        \toprule
        \textbf{SYSTEM:} You are an expert scientific source identifier. You will be given a social media post that paraphrases a scientific finding, and a candidate scientific paper. Your job is to judge how likely the post is referring to this specific paper. Output ONLY a single integer score from 0 to 10 (no explanation, no other text). 10 = the post is almost certainly about this paper. 0 = the paper is unrelated. \\
        \midrule
        \textbf{USER:} Social media post: \{post\} \newline Candidate paper: \newline Title: \{title\} \newline Abstract: \{abstract\} \newline Score (0-10): \\
        \bottomrule
    \end{tabular}
\end{table}

\subsection{Implementation Details}

We ran all experiments on a single node with two NVIDIA A40 GPUs (46~GB each) and 128~GB of memory. Models are loaded via Hugging Face transformers, dense encoding uses sentence-transformers, and BM25 runs on a pre-tokenized corpus. We serve the LLM reranker with vLLM for high-throughput inference. End-to-end inference on the English test set (303{,}800 prompts) takes roughly 11 hours.

\section{Result and Analysis}

\begin{table*}[t]
    \centering
    \caption{English dev and test set results. All scores against the gold pubkey labels; best score per column in \textbf{bold}.}
    \label{tab:results}
    \resizebox{\textwidth}{!}{%
    \begin{tabular}{@{}cllcccccc@{}}
        \toprule
         & & & \multicolumn{3}{c}{\textbf{Dev ($N = 3{,}905$)}} & \multicolumn{3}{c}{\textbf{Test ($N = 6{,}076$)}} \\
        \cmidrule(lr){4-6} \cmidrule(lr){7-9}
        \# & System & Stage & MRR@5 & Hit@1 & Hit@5 & MRR@5 & Hit@1 & Hit@5 \\
        \midrule
        1 & BM25 & Lexical & 0.5069 & 0.4584 & 0.5828 & 0.5021 & 0.4493 & 0.5859 \\
        2 & E5-large-instruct (zero-shot) & Dense & 0.5265 & 0.4638 & 0.6292 & 0.5164 & 0.4536 & 0.6177 \\
        3 & BM25 $\oplus$ E5 (RRF, $k=60$) & Hybrid & 0.5419 & 0.4761 & 0.6487 & 0.5577 & 0.4924 & 0.6587 \\
        4 & Hybrid + Cross-encoder rerank (bge-reranker-v2-m3, top-30) & Hybrid + CE & 0.5512 & 0.4912 & 0.6423 & 0.5422 & 0.4724 & 0.6513 \\
        5 & BM25 + Qwen2.5-14B rerank & Lexical + LLM & 0.6073 & 0.5649 & 0.6676 & 0.6062 & 0.5566 & 0.6746 \\
        6 & Hybrid + Qwen2.5-14B rerank (submitted) & Hybrid + LLM & \textbf{0.6436} & \textbf{0.5836} & \textbf{0.7347} & \textbf{0.6439} & \textbf{0.5797} & \textbf{0.7355} \\
        \bottomrule
    \end{tabular}%
    }
\end{table*}

We follow the official CLEF 2026 CheckThat! Lab Task 1 protocol and use \textbf{Mean Reciprocal Rank (MRR)} as our primary evaluation metric:

\begin{equation}
\mathrm{MRR} = \frac{1}{|Q|} \sum_{i=1}^{|Q|} \frac{1}{\mathrm{rank}_i}
\end{equation}

where $|Q|$ is the number of queries in the evaluation set, and $\mathrm{rank}_i$ is the rank position of the gold-standard paper for the $i$-th post within the top-5 predictions (taken as $0$ if the gold paper is not retrieved within the top 5). MRR is well suited to this task, as it rewards systems that rank correct evidence near the top of the list.

We evaluate all systems on the English development split (3{,}905 posts) and the official English test split (6{,}076 posts). In addition to MRR@5; we report Hit@1 and Hit@5 to characterize where in the ranking the correct paper is placed. Table~\ref{tab:results} summarizes the contribution of each component of the pipeline.

Our result analysis showed by using lexical retrieval methods like BM25 both dev set and test set achieves MRR@5 of 50\%. The near-identical scores across splits indicate that the corpus and query distributions are well matched, and establish a stable reference point against which the subsequent retrieval stages are evaluated. On the other hand semantic retrieval like the multilingual encoder \texttt{intfloat/multilingual-e5-large-instruct} in a zero-shot configuration yields an absolute improvement of 1.96 percentage points in MRR@5 over the BM25 baseline, reaching 52.65\% on the development set, with a substantially larger 4.64-point gain in Hit@5 (58.28\% to 62.92\%). This confirms that the instruction-tuned dense encoder captures the implicit semantic links between social-media claims and scientific abstracts more reliably than surface term matching. Combining the BM25 and zero-shot dense rankings with Reciprocal Rank Fusion provides a further +1.54 percentage points in MRR@5 over the dense leg alone (54.19\% vs.\ 52.65\%) and +1.95 in Hit@5, confirming that the two signals are complementary.

In subsequent stage we compare two re-rankers operating on the hybrid fused candidates: a cross-encoder over the top-30, and a generative LLM re-ranker over the top-50. The cross-encoder \texttt{BAAI/bge-reranker-v2-m3} contributes a modest +0.93-point gain in MRR@5 over the hybrid first stage (55.12\% vs.\ 54.19\%), with most of the benefit concentrated at the top of the ranking (Hit@1 improves from 47.61\% to 49.12\%). However, this gain reverses on the test split, where the cross-encoder loses 1.55 points relative to the hybrid baseline (54.22\% vs.\ 55.77\%)---indicating that cross-encoder reranking overfits to the development distribution. The 14B-parameter generative re-ranker Qwen2.5-14B-Instruct delivers a substantially larger improvement: on the development set, applying it on top of the hybrid candidates raises MRR@5 by +10.17 percentage points (54.19\% to 64.36\%) and Hit@5 by +8.60 points (64.87\% to 73.47\%). Applied directly to BM25 candidates, the same re-ranker yields an MRR@5 of 60.73\%, indicating that approximately three-quarters of its benefit is independent of the first-stage retriever.

Our final pipeline with LLM re-ranking over the hybrid top-50 achieves an MRR@5 of 64.36\% on the development set and 64.39\% on the test set, with corresponding Hit@1 of 58.36\% and Hit@5 of 73.47\% and the test (57.97\% and 73.55\%) respectively. The minimal 0.03-point gap between development and test MRR@5 confirms that the tuning decisions transfer cleanly to held-out data; the LLM re-ranker is the dominant contributor to the +13.67-point overall improvement in MRR@5 over the BM25 baseline.

\section{Discussion \& Conclusion}

Two findings strengthen our empirical contributions to literature.

\textbf{The LLM reranker drives the gain.} Qwen2.5-14B contributes +10 MRR@5 points whether applied to BM25 or hybrid candidates roughly three-quarters of the +13.67-point overall lift over BM25. Pointwise scoring with a strong open-weight LLM is a reliable substitute for training a task-specific reranker~\cite{sun2024}.

\textbf{Hybrid retrieval compounds with reranking.} RRF fusion yields a modest +1.54-point first-stage gain but a larger +3.63-point lift after LLM reranking (Hybrid+LLM 64.36\% vs.\ BM25+LLM 60.73\%): better first-stage recall translates directly into stronger downstream rankings~\cite{bruch2023,yang2019}.

Our full pipeline is open-source, modular, and runs end-to-end on a single two-GPU node with vLLM serving make our pipeline practical for privacy-sensitive or offline deployment.

\section{Limitations \& Future Work}

Our set up was evaluated on the English split only and represent each paper by title and abstract, as a result the German and French test splits and metadata-aware representations are left for future work. The Qwen2.5-14B reranker is also the runtime bottleneck ($\sim$11 hours per 6{,}076 test queries on two A40 GPUs). Future directions will include (i) Retrieval Augmented Generation (RAG) along with the metadata feature feeding into the LLM for additional cues for guided ranking (ii) distilling the LLM reranker into a smaller cross-encoder or adopting listwise inference to reduce per-query cost.

\section*{Declaration on Generative AI}

During the creation of this work, the authors used Claude to refine the pre-written text and to create Figure 1. After using the tool, the authors reviewed and edited the content as needed and take full responsibility for the publication's content.

\bibliography{sample-ceur}

\end{document}